\documentstyle[12pt]{article}

\input tcilatex
\QQQ{Language}{
American English
}

\begin{document}

\title{QCD One-Loop effective coupling constant and effective quark mass given in a
mass-dependent renormalization }
\author{Jun-Chen Su, Lian-You Shan and Ying-Hui Cao \\
Center for Theoretical Physics, Department of Physics,\\
Jilin University, Changchun 130023,\\
People's Republic of China}
\date{}
\maketitle

\begin{abstract}
The QCD one-loop renormalization is restudied in a mass-dependent
subtraction scheme in which the quark mass is not set to vanish and the
renormalization point is chosen to be an arbitrary time-like momentum. The
correctness of the subtraction is ensured by the Ward identities which are
respected in all the processes of subtraction. By considering the mass
effect, the effective coupling constant and the effective quark mass are
given in improved expressions which are different from the previous results.

PACS: 11.10Gh, 12.38.Bx

Keywords: QCD renormalization, mass-dependent subtraction, time-like
renormalization point, effective coupling constant and quark mass, exact
one-loop results.
\end{abstract}

\setcounter{section}{1}

\section*{1.Introduction}

~~~The asymptotically free property of Quantum Chromodynamics (QCD) has been
widely applied to analyze the scaling behaviors of high energy processes and
attracted much attention to investigate the QCD renormalization$^{1-23}$.
This property was at first discovered in Refs.(1) and (2) from studying the
one-loop renormalization and the solution of renormalization group equation
(RGE)$^{24-29}$satisfied by the coupling constant. The anomalous dimension (
or the $\beta $-function) appearing in the equation was computed from the
renormalization constants which are ordinarily given by mass-independent
subtractions such as the minimal subtraction (MS)$^6$, the modified minimal
subtraction$^7$and the momentum space subtraction(MOM)$^8$ . However, there
appears a serious ambiguity problem$^{11}$ that different subtraction
schemes give different results for a renormalized quantity in a finite order
perturbative calculation. In the $MS$ schemes which is carried out in the
dimensional regularization procedure, only the divergent terms proportional
to the factor $\frac 1\varepsilon $ (where $\varepsilon =2-\frac n2$) are
subtracted. The $MS$ scheme was demonstrated to give the perturbative
expansion in the coupling constant for a physical quantity which is of worse
convergence. To improve the convergence, the $\overline{MS}$ scheme was
proposed to subtract the divergent part of a Feynman diagram which contains
the factor $\frac 1{\widehat{\varepsilon }}=\frac 1\varepsilon +\ln 4\pi
-\gamma $ where the unphysical terms $\ln 4\pi -\gamma $ arise from a
special way of analytical continuation of the dimension from n to 4$^9$. The 
$MS$ and $\overline{MS}$ schemes are simple and respect the Ward identity,
but considered to be nonphysical$^8$. In the MOM scheme, the divergence in a
Feynman integral is subtracted at an Euclidean momentum, $p_i^2=-\mu ^2$.
This subtraction was viewed as physical and shown to give the perturbative
series of better convergence. Nevertheless, the renormalization constants
obtained in this scheme do not automatically satisfy the Ward identity. To
solve the ambiguity problem, several prescriptions were proposed in the past$%
^{10-14}.$ In our preceding paper on the QED renormalization$^{33}$, we
treated the ambiguity problem from a different angle. It was argued that a
subtraction scheme must respect necessary physical and mathematical
principles shcu as the gauge-symmetry (the Ward identity), the
Lorentz-invariance( the energy-momentum conservation) and the mathematical
convergence. In addition, it was proved that the solution of the
renormalization group equation (RGE) for a renormalized quantity (for
example, a wave function, a propagator or a vertex) can be uniquely
determined by applying the renormalization boudary condition without any
ambiguity. As a result, an exact S-matrix element can be expressed in the
form as given in the tree-diagram approximation except that the coupling
constant, the fermion mass and the gauge parameter become  effective
(running )ones. Therefore the tesk of renormalization is reduced to find
these effective quantities.

In the mass-independent subtractions mentioned before, the fermion mass is
set to be zero, or say, only the massless fermion propagator $\frac 1{\not p}
$ is used in the calculation of divergent Feynman diagrams. The rationality
of the mass-independent subtraction was argued as follows$^{6,29}$. In the
conventional perturbation series which is expanded in coupling constant, the
massive fermion propagator can be expanded as such a series

\begin{equation}
\frac 1{\not p-m}=\frac 1{\not p}+\frac 1{\not p}m\frac 1{\not p}+\frac 1{%
\not p}m\frac 1{\not p}m\frac 1{\not p}+\cdot \cdot \cdot \cdot \cdot \cdot
\end{equation}
Substitution of this series into the coventional perturbation series leads
to a new perturbation series in which the fermion propagator becomes
massless and the fermion mass, like the coupling constant, can be treated as
another expansion parameter. Working with this kind of perturbation series,
one may perform the mass-independent subtraction. It is obvious that in
order to obtain a quantity which is of a certain order of coupling constant
in the sense of the conventional perturbation theory, according to Eq.(1.1),
one has to compute an infinite set of perturbative terms in the new
perturbation theory. In the large momentum limit, the massive propagator is
reduced to the massless one. In this case, there will be no difference
between the both perturbation theories mentioned above. However, for a
process whose energy is not too high, particularly, in the energy region
near the meson production threshold, the mass in the fermion propagator can
not be negligible. In this case, to see the effect of fermion mass on the
renormalization , it is convenient to carry out a mass-dependent
renormalization in the framework of the conventional perturbation theory as
was done in the previous literature$^{3,10}$. Obviously, the MOM scheme is
suitable for this kind of renormalization. In Ref.(3), the subtraction was
carried out at an arbitrary Euclidean point, $p_i^2=-\mu ^2$ and in the
Landau gauge. while, in Ref.(10), the renormalization was done at an off
mass shell point, $p_i^2-m_i^2=-\mu ^2$ and in the Feynman gauge. Both of
the subtractions give the same one-loop anomalous dimension of the coupling
constant, but different anomalous dimensions of the quark mass.

In this paper, we wish to restudy the QCD mass-dependent renormalization in
the MOM\ scheme along the line as described in our preceding paper.  The new
features of this study are: (1) The subtraction exactly respects necessary
physical and mathematical principles. Owing to the restriction of these
principles, the renormalization will be put on the faithful basis and have
no ambiguity. For example, by the convergence principle, we are not allowed
to employ the divergent form of renormalization constants to do a meaningful
calculation because a divergent quantity is not well-defined mathematically
and therefore can not be unambiguously evaluated by any computational rule.
The correct procedure of computing the anomalous dimension in a RGE is usage
of regularized froms of the renormalization constants. By this procedure, a
renormalized quantity given by solving its RGE will be uniquely determined
by its anomalous dimension and boundary condition; (2) the renormalization
point is taken to be a time-like (Minkowski) momentum. The subtraction with
this renormalization point will be called generalized mass-shell scheme(GMS)
because this scheme can naturally lead to the results given in the
mass-shell subtraction scheme. It will be shown that the behavior of a
renormalized quantity derived in the GMS scheme is different from the ones
obtained in other subtraction schemes; (3) The QCD one-loop effective
coupling constant and effective quark masses derived in the GMS scheme and
in general gauges are given rigorous and explicit expressions. These
expressions will go over to the results given in the MS scheme in the large
momentum limit.

The remainder of this paper is arranged as follows. In Sect.2, we will
describe the derivation of the renormalization constant of coupling
constant. This renormalization constant is chosen to be determined by the
renormalization constants of the gluon propagator, the ghost particle
propagator and the ghost vertices. The renormalization constant of gluon
propagator is given based on the Ward identity. It will be shown that the
other two renormalization constants for the ghost particle propagator and
the ghost vertex will give their anomalous dimensions which are
renormalization-scheme-independent. In Sect.3, we will derive an explicit
expression of the effective coupling constant by solving its RGE and show
its asymptotic behavior. In Sect.4, we will give a derivation of the Ward
identity satisfied by the quark-gluon vertex and prove that in the
approximation of order $g^2$, this identity is reduced to the form as found
in QED and thus the subtraction version of the quark self-energy is similar
to that for the electron self-energy. In Sect.5, an explicit expression of
effective quark masses will be given by solving its RGE. We will end this
paper with some comments and discussions.

\setcounter{section}{2}

\section*{2. The Renomalization Constant of Coupling Constant}

\setcounter{equation}{0}

The renormalization constant of coupling constant $Z_g$ is defined by 
\begin{equation}
g=Z_gg_R
\end{equation}
where $g_R$ and $g$ denote the renormalized and unrenormalized coupling
constants respectively. According to the Ward identity$^{30}$, the $Z_g$ can
be expressed in different ways. In this section, we prefer to use the
following expression 
\begin{equation}
Z_g=\frac{\widetilde{Z}_1}{\widetilde{Z}_3Z_3^{\frac 12}}
\end{equation}
where $Z_3,\widetilde{Z}_3$ and $\widetilde{Z}_1$ are the renormalization
constants for the gluon propagator, the ghost particle propagator and ghost
vertex respectively. In the following, we will describe how they are
determined by the subtractions of the gluon self-energy, the ghost particle
self-energy and the ghost vertex correction in the GMS scheme. The one-loop
diagrams for the gluon self-energy, the ghost particle self-energy and the
ghost vertex correction have already been calculated in the literature$%
^{8,31,32}$ by the dimensional regularization . So, we may directly quote
the results and put some emphases on essential points of the subtraction
procedure in the GMS scheme.

~~The renormalization constant $Z_3$ is, in the GMS scheme, defined by 
\begin{equation}
Z_3^{-1}=1+\Pi (\mu ^2)
\end{equation}
where $\Pi (k^2)$ is the gluon self-energy appearing in the transverse part
of gluon propagator 
\begin{equation}
iD_{\mu \nu }^{ab}(k)=\frac{-i\delta ^{ab}}{k^2+i\varepsilon }[(g_{\mu \nu }-%
\frac{k_\mu k_\nu }{k^2})\frac 1{1+\Pi (k^2)}+\xi \frac{k_\mu k_\nu }{k^2}]
\end{equation}
A correct way of calculating the function $\Pi (k^2)$ is to use the relation 
\begin{equation}
\Pi (k^2)=\frac 1{3k^2}g^{\mu \nu }\Pi _{\mu \nu }(k)
\end{equation}
where $\Pi _{\mu \nu }(k)$ is the radiative correction tensor which is
transverse, $\Pi _{\mu \nu }(k)=(k^2g_{\mu \nu }-k_\mu k_\nu )\Pi (k^2)$, as
implied by the Ward identity $k^\mu \Pi _{\mu \nu }(k)=0$. For the one-loop
gluon self-energy diagrams as depicted in Figs.(1a)-(1c), after computing
the $\Pi (k^2)$ by the dimensional regularization procedure and then setting 
$k^2=\mu ^2$, one can get the regularized form of the $Z_3$ as follows 
\begin{eqnarray}
Z_3 &=&1+\frac{g^2}{(4\pi )^2}(4\pi M^2)^\varepsilon \Gamma (1+\varepsilon
)\int_0^1dx\{\frac 1{[\mu ^2x(x-1)]^\varepsilon }  \nonumber \\
&&\times [\frac 1\varepsilon J_1(x)+J_2(x)]+\frac 1\varepsilon
\sum_{i=1}^{N_f}\frac{J_3(x)}{[\mu ^2x(x-1)+m_i^2]^\varepsilon }\}
\end{eqnarray}
where 
\begin{eqnarray}
J_1(x) &=&2(n-1)x(x-1)\{\frac 1{n-2}[3n-4+2(1-\xi )(x-1)]  \nonumber \\
&&+\frac 3{2x(x-1)}+2(1-\xi )[2(1+\frac 1n)x-(3+\frac 4n)+\frac 3{nx}]\}
\end{eqnarray}
\begin{eqnarray}
J_2(x) &=&-(1-\xi )(n-1)[4(1+\frac 1n)x(x-1)(x-2)+(1+\frac 6n)x  \nonumber \\
&&-\frac 6n-\frac 14(1-\xi )]
\end{eqnarray}
\begin{equation}
J_3(x)=\frac 13n(n-1)x(x-1)
\end{equation}
and $m_i$ is the mass of i-th quark, while, M is an arbitrary mass
introduced to make the coupling constant g dimensionless in the
n-dimensional space.

~~The constant $\widetilde{Z}_3$ is, in the GMS scheme, defined as 
\begin{equation}
\widetilde{Z}_3^{-1}=1+\omega (\mu ^2)
\end{equation}
where $\omega (q^2)$ is the self-energy of ghost particle which appears in
the ghost particle propagator 
\begin{equation}
i\Delta ^{ab}(q)=\frac{-i\delta ^{ab}}{q^2[1+\omega (q^2)]+i\varepsilon }
\end{equation}
For the one-loop diagram shown in Fig.(1e), the regularized form of the $%
\widetilde{Z}_3$ is easily obtained by the dimensional regularization as
shown in the following 
\begin{eqnarray}
\widetilde{Z}_3 &=&1+\frac{g^2}{(4\pi )^2}(4\pi M^2)^\varepsilon \Gamma
(1+\varepsilon )\int_0^1dx3(1-x)\{{\frac 1{\varepsilon [\mu
^2x(x-1)]^\varepsilon }}  \nonumber \\
{}{} &&{\times [\frac 12(1+\xi )+3(1-\xi )x]+\frac 14(1-\xi )(1-10x)\}}
\end{eqnarray}

Now, let us turn to discuss the subtraction of the ghost vertex. The vertex
is generally represented as 
\begin{equation}
\Gamma _\mu ^{abc}(p,q)=gf^{abc}[p_\mu +\Lambda _\mu (p,q)]
\end{equation}
where $\Lambda _\mu (p,q)$ denotes the higher order corrections. According
to the Lorentz covariance, it may be written in the form 
\begin{equation}
\Lambda _\mu (p,q)=A(p^2,q^2,p\cdot q)p_\mu +B(p^2,q^2,p\cdot q)q_\mu
\end{equation}
where A and B are the scalar functions. Suppose only the function A is
divergent in the limit $\varepsilon \rightarrow 0$, while the B is finite as
we encountered in the one-loop approximation. In this case, we may only
subtract the divergent part in the A at the point $p^2=q^2=\mu ^2$.
Alternatively, it is more convenient to choose the renormalization point
such that p=q and $p^2=\mu ^2$ which implies $r^2=(p-q)^2=0$. This choice is
compatible with the energy-momentum conservation and yields 
\begin{equation}
\Lambda _\mu (p,q)\mid _{p=q,p^2=\mu ^2}=\widetilde{L}p_\mu
\end{equation}
where 
\begin{equation}
\widetilde{L}=A(\mu ^2)+B(\mu ^2)
\end{equation}
With the subtraction in Eq.(2.15) and the definition 
\begin{equation}
\widetilde{Z}_1^{-1}=1+\widetilde{L}
\end{equation}
The vertex in Eq.(2.13) may be renormalized as 
\begin{equation}
\Gamma _\mu ^{abc}(p,q)=\widetilde{Z}_1^{-1}\Gamma _{R\mu }^{abc}(p,q)
\end{equation}
where 
\begin{equation}
\Gamma _{R\mu }^{abc}(p,q)=gf^{abc}[p_\mu +\Lambda _{R\mu }(p,q)]
\end{equation}
which is the renormalized vertex. The latter vertex satisfies the boundary
condition 
\begin{equation}
\Gamma _{R\mu }^{abc}(p,q)\mid _{p=q,p^2=\mu ^2}=gf^{abc}p_\mu
\end{equation}
which is the bare vertex, showing the advantage of the subtraction chosen.
For the one-loop diagrams depicted in Figs.(1f) and (1g), according to the
definition in Eqs.(2.15) and (2.16) and through tedious calculations, we
find in the n-dimensional space 
\begin{equation}
\widetilde{L}=\widetilde{L}_1+\widetilde{L}_2
\end{equation}
where $\widetilde{L}_1$ and $\widetilde{L}_2$ are given by diagrams in
Figs.(1h) and (1i) respectively. 
\begin{equation}
\widetilde{L}_1=\frac{g^2}{(4\pi )^2}(4\pi M^2)^\varepsilon \Gamma
(1+\varepsilon )\int_0^1dy{\frac{3y}{\varepsilon [\mu ^2y(y-1)]^\varepsilon }%
\frac 1n[3\xi -2+3(1-\xi )y]+fts}
\end{equation}
and 
\begin{eqnarray}
\widetilde{L}_2 &=&\frac{g^2}{(4\pi )^2}(4\pi M^2)^\varepsilon \Gamma
(1+\varepsilon )\int_0^1dy{\frac{3y}{\varepsilon [\mu ^2y(y-1)]^\varepsilon }%
\frac{(n-1)}n[1-\frac 32(1-\xi )y]}  \nonumber \\
&&{+fts}
\end{eqnarray}
where the symbol''fts'' represents the terms which are finite in the limit $%
\varepsilon \to 0$. These terms are not necessary to be written explicitly
for our purpose because they are independent of the renormalization point $%
\mu $ and therefore give no contributions to the anomalous dimension. In the
approximation of order $g^2$, considering Eq.(2.21), the $\widetilde{Z}_1$
defined in Eq.(2.17) can be written as 
\begin{equation}
\widetilde{Z}_1=1-\widetilde{L}_1-\widetilde{L}_2
\end{equation}

\setcounter{section}{3}

\section*{3.Effective Coupling Constant}

\setcounter{equation}{0}

~~~The effective coupling constant is determined by the following RGE 
\begin{equation}
\mu \frac d{d\mu }g_R(\mu )+\gamma _g(\mu )g_R(\mu )=0
\end{equation}
which is obtained by differentiating Eq.(2.1) with respect to $\mu $.
According to the definition in Eq.(2.2), the anomalous dimension $\gamma
_g(\mu )$ is given by 
\begin{equation}
\gamma _g=\lim_{\varepsilon \to 0}\mu \frac d{d\mu }\ln Z_g=\widetilde{%
\gamma }_1-\widetilde{\gamma }_3-\frac 12\gamma _3
\end{equation}
The anomalous dimensions $\gamma _3,\widetilde{\gamma }_3$ and $\widetilde{%
\gamma }_1$ in the approximation of order $g^2$ are easily calculated from
the corresponding renormalization constants. From Eqs.(2.6)-(2.9), we obtain 
\begin{eqnarray}
\gamma _3 &=&\lim_{n\rightarrow 4}\mu \frac d{d\mu }\ln Z_3(\mu ,n) 
\nonumber \\
\ &=&\frac{g^2}{(4\pi )^2}\{3\xi -13+\frac 43\sum_{i=1}^{N_f}[1+6\sigma _i^2+%
\frac{12\sigma _i^4}{\sqrt{1-4\sigma _i^2}}  \nonumber \\
&&\ \times \ln \frac{1+\sqrt{1-4\sigma _i^2}}{1-\sqrt{1-4\sigma _i^2}}]\}
\end{eqnarray}
where $\sigma _i=\frac{m_R^i}\mu $. From Eq.(2.12), it follows 
\begin{equation}
\widetilde{\gamma }_3=\lim_{n\rightarrow 4}\mu \frac d{d\mu }\ln \widetilde{Z%
}_3(\mu ,n)=\frac{g^2}{(4\pi )^2}(\frac 32\xi -\frac 92)
\end{equation}
By using the expressions given in Eqs.(2.21)-(2.24), it is easy to find 
\begin{equation}
\widetilde{\gamma }_1=\lim_{n\rightarrow 4}\mu \frac d{d\mu }\ln \widetilde{Z%
}_1(\mu ,n)=\frac{g^2}{(4\pi )^2}3\xi
\end{equation}
There are three points we would like to stress here. (1) The results given
in Eqs.(3.4) and (3.5) are exactly identical to those obtained in the MS
scheme. This is because in the one-loop diagrams of the ghost particle
self-energy and the ghost vertex, only the massless particles are involved.
So, the $\widetilde{\gamma }_1$ and $\widetilde{\gamma }_3$ given above are
scheme-independent. This is why we like to choose the ghost particle
self-energy and the ghost vertex to define the renormalization constant $Z_g$
. As for the $\gamma _3$, it is noted that the term related to the $J_2(x)$
in Eq.(2.6) which is finite in the limit$\varepsilon \rightarrow 0$ gives no
contribution to the anomalous dimension owing to its independence of the
fermion mass. When the quark mass is set to be zero, the result in Eq.(3.3)
will be reduced to that given in the MS scheme. In this case, the $\gamma _3$
is only given by the terms in Eq.(2.6) which contain the $\varepsilon $%
-pole. (2) Each of the quark bare masses in Eq.(3.3) has been replaced by
its renormalized one which is taken to be a constant and identified with the
pole of the quark propagator. The replacement is suitable only in the lowest
order approximation. At one-loop level, we concern the anomalous dimension
of the order of $g^2$, therefore, the quark mass in Eq.(3.3) is only needed
to be given in the lowest order. (3) The renormalization point may be
parametrized by a scale variable $\lambda ,\mu =\mu _0\lambda $ where $\mu
_0 $ is a fixed scale parameter which may be chosen from physical
consideration. The $\lambda $ can also be chosen to be the scale variable
for momenta, $p=\lambda p_0$. With the $\lambda $ introduced, we may write $%
\sigma _i=\frac{C_i}\lambda $ where $c_i=\frac{m_R^i}{\mu _0}$. Noticing
this relation, on inserting Eqs.(3.3)-(3.5) into Eq.(3.2), the anomalous
dimension $\gamma _g$ will be represented in the following 
\begin{equation}
\gamma _g(\lambda )=\frac{g_R^2}{(4\pi )^2}F_g(\lambda )
\end{equation}
where 
\begin{equation}
F_g(\lambda )=11-\frac 23\sum_{i=1}^{N_f}[1+\frac{6c_i^2}{\lambda ^2}+\frac{%
12c_i^4}{\lambda ^3}f_i(\lambda )]
\end{equation}
in which 
\begin{eqnarray}
f_i(\lambda ) &=&\frac 1{\sqrt{\lambda ^2-4c_i^2}}\ln \frac{\lambda +\sqrt{%
\lambda ^2-4c_i^2}}{\lambda -\sqrt{\lambda ^2-4c_i^2}}  \nonumber \\
\ &=&\cases{ \frac 2{\sqrt{4c_i^2-\lambda ^2}}\cot ^{-1}\frac \lambda
{\sqrt{4c_i^2-\lambda ^2}},&if $ \lambda \leq 2c_i$ \cr \frac
2{\sqrt{\lambda ^2-4c_i^2}}\coth ^{-1}\frac \lambda {\sqrt{\lambda
^2-4c_i^2}},&if $\lambda \geq 2c_i$ \cr}
\end{eqnarray}
Substituting Eq.(3.6) into Eq.(3.1) and noticing $\mu \frac d{d\mu }=\lambda
\frac d{d\lambda }$, the equation may be written as 
\begin{equation}
\frac{dg_R}{g_R^3}=-\frac 1{(4\pi )^2}F_g(\lambda )\frac{d\lambda }\lambda
\end{equation}
On integrating the above equation by applying the familiar integration
formulas, we obtain 
\begin{equation}
\alpha _R(\lambda )=\frac{\alpha _R^0}{1+\frac{\alpha _R^0}{2\pi }G(\lambda )%
}
\end{equation}
where $\alpha _R=\frac{g_R^2}{4\pi }$ and 
\begin{equation}
G(\lambda )=\int_1^\lambda \frac{d\lambda }\lambda F_g(\lambda )=11\ln
\lambda -\frac 23[a-\frac{2b}{\lambda ^2}+\varphi (\lambda )]
\end{equation}
in which 
\begin{equation}
a=\sum_{i=1}^{N_f}[2c_i^2-(1+2c_i^2)\chi _i]
\end{equation}
here 
\begin{eqnarray}
\chi _i &=&\sqrt{1-4c_i^2}\ln \frac 1{2c_i}(1+\sqrt{1-4c_i^2})  \nonumber \\
\ &=&\cases{ -\sqrt{4c_i^2-1}\cos ^{-1}\frac 1{2c_i},&if $ 2c_{i\geq 1}$\cr
\sqrt{1-4c_i^2}\cosh ^{-1}\frac 1{2c_i},&if $ 2c_{i\leq 1}$\cr}
\end{eqnarray}
\begin{equation}
b=\sum_{i=1}^{N_f}c_i^2
\end{equation}
and 
\begin{equation}
\varphi (\lambda )=\sum_{i=1}^{N_f}(1+\frac{2c_i^2}{\lambda ^2})\frac
1\lambda \eta _i(\lambda )
\end{equation}
here 
\begin{eqnarray}
\eta _i(\lambda ) &=&\sqrt{\lambda ^2-4c_i^2}\ln \frac 1{2c_i}(\lambda +%
\sqrt{\lambda ^2-4c_i^2})  \nonumber \\
\ &=&\cases{ -\sqrt{4c_i^2-\lambda ^2}\cos ^{-1}\frac \lambda {2c_i},&if $
\lambda \leq 2c_i$ \cr \sqrt{\lambda ^2-4c_i^2}\cosh ^{-1}\frac \lambda
{2c_i},&if $ \lambda \geq 2c_i$\cr}
\end{eqnarray}

If we set all the quark masses to be equal and choose $\mu _0=m_R$,
Eqs.(3.7) and (3.11) will be respectively reduced to 
\begin{equation}
F_g(\lambda )=11-\frac 23N_f[1+\frac 6{\lambda ^2}+\frac{12}{\lambda ^3\sqrt{%
\lambda ^2-4}}\ln \frac{\lambda +\sqrt{\lambda ^2-4}}{\lambda -\sqrt{\lambda
^2-4}}]
\end{equation}
and 
\begin{equation}
G(\lambda )=11\ln \lambda -\frac 23N_f[2+\sqrt{3}\pi -\frac 2{\lambda
^2}+(1+\frac 2{\lambda ^2})\frac 1\lambda \eta (\lambda )]
\end{equation}
where 
\begin{eqnarray}
\eta (\lambda ) &=&\sqrt{\lambda ^2-4}\ln \frac 12(\lambda +\sqrt{\lambda
^2-4})  \nonumber \\
\ &=&\cases{ -\sqrt{4-\lambda ^2}\cos ^{-1}\frac \lambda 2,&if $\lambda \leq
2$ \cr \sqrt{\lambda ^2-4}\cosh ^{-1}\frac \lambda 2,&if $ \lambda \geq
2$\cr}
\end{eqnarray}
In the above formulas, the $\lambda $ may be defined by $\lambda =\left|
q^2/m_R^2\right| ^{\frac 12}$. When $\lambda =1$, due to G(1)=0, Eq.(3.10)
becomes $\alpha _R(1)=\alpha _R^0$. This gives the result on the fermion
mass shell. When $\lambda \rightarrow \infty $, i.e. in the large momentum
limit. it is easy to see from Eqs.(3.18) and (3.19) that $G(\lambda
)\rightarrow (11-\frac 23N_f)\ln \lambda $. Thus, Eq.(3.10) is reduced to 
\begin{equation}
\alpha _R(\lambda )=\frac{\alpha _R^0}{1+\frac{\alpha _R^0}{6\pi }%
(33-2N_f)\ln \lambda }
\end{equation}
This just is the result obtained previously in the MS scheme$^{1,2}$. The
behavior of the effective coupling constants are represented in Figs.(2) and
(3). The coupling constants in Fig.(2) are given by taking all the quark
masses to be equal and $N_f=4$. The solid curve represents the coupling
constant evaluated by using Eqs.(3.10), (3.18) and (3.19). Fig.(3) shows
some coupling constants for different numbers of flavor and unequal masses.
In the figure, the constituent quark masses are taken to be $%
m_u=m_d=0.3GeV,m_s=0.45GeV,m_c=1.5GeV,m_b=5.0GeV$ and $m_t=175GeV$, and the
scale $\mu _0$ is taken to be equal to the $N_f$-th quark mass. From
Figs.(2) and (3), one can see that all the $\alpha _R(\lambda )$ decrease
with the increase of $\lambda $ and tend to zero when $\lambda \rightarrow
\infty $, exhibiting the well-known asymptotically free property. In the
region near $\lambda =1$, there is a maximum for each curve. The height and
the position of the maximum weakly depend on the number of flavor(see
Fig.(3)). In comparison with the result given in the MS scheme, as we see
from Fig.(2), the quark mass gives a considerable improvement on the
effective interaction, particularly, in the region near the heavier meson
threshold. However, as shown in Figs.(2) and (3). When $\lambda \rightarrow
0 $, all the $\alpha _R(\lambda )$ drastically fall down to zero from their
maxima. This unreasonable result indicates that the QCD perturbation theory
is inapplicable in the very small momentum domain.

\setcounter{section}{4}

\section*{4.The Ward Identity for Quark-Gluon Vertex}

\setcounter{equation}{0}

~~~The aim of this section is to sketch the derivation of the Ward identity
satisfied by the quark-gluon vertex and to show how we should do for the
subtraction of the quark self-energy. The Ward identity is not difficult to
derive from the QCD generating functional. Firstly, we write the Ward
identity obeyed by the quark-gluon three point Green function$^{32}$ 
\begin{eqnarray}
\partial _z^\mu &<&0^{+}\mid T[\widehat{\psi }(x)\widehat{\overline{\psi }}%
(y)\widehat{A}_\mu ^a(z)]\mid 0^{-}>  \nonumber \\
&=&i\xi g\{<0^{+}\mid T[\widehat{\psi }(x)\overline{\psi }(y)\widehat{C}^b(y)%
\widehat{\overline{C}}^a(z)]\mid 0^{-}>T^b  \nonumber \\
-T^b &<&0^{+}\mid T[\widehat{\psi }(x)\widehat{\overline{\psi }}(y)\widehat{C%
}^b(x)\widehat{\overline{C}}^a(z)]\mid 0^{-}>\}
\end{eqnarray}
where $T^b$ denote the generators of SU(3) group, $\widehat{\psi }(x),%
\widehat{A}_\mu ^a(x)$ and $\widehat{C}^a(x)$ represent the field operators
for quark, gluon and ghost particle respectively. The Green's functions in
Eq.(4.1) have the following irreducible decompositions$^{31,32}$ 
\begin{eqnarray}
&<&0^{+}\mid T[\widehat{\psi }(x)\widehat{\overline{\psi }}(y)\widehat{A}%
_\mu ^a(z)]\mid 0^{-}>  \nonumber \\
&=&\int d^4x^{\prime }d^4y^{\prime }d^4z^{\prime }S_F(x-x^{\prime })\Gamma
^{b\nu }(x^{\prime },y^{\prime },z^{\prime })S_F(y^{\prime }-y)D_{\nu \mu
}^{ba}(z^{\prime }-z)
\end{eqnarray}
\begin{eqnarray}
&<&0^{+}\mid T[\widehat{\psi }(x)\widehat{\overline{\psi }}(y))\widehat{C}%
^b(y)\widehat{\overline{C}}^a(z)]\mid 0^{-}>  \nonumber \\
&=&\int d^4x^{\prime }d^4z^{\prime }S_F(x-x^{\prime })\gamma ^{ba^{\prime
}}(x^{\prime },y,z^{\prime })\Delta ^{aa^{\prime }}(z-z^{\prime })
\end{eqnarray}
where 
\begin{equation}
\gamma ^{ba^{\prime }}(x^{\prime },y,z^{\prime })=\int d^4y^{\prime
}d^4u^{\prime }\Gamma ^{b^{\prime }a^{\prime }}(x^{\prime },y^{\prime
},u^{\prime },z^{\prime })S_F(y^{\prime }-y)\Delta ^{b^{\prime }b}(u^{\prime
}-y)
\end{equation}
and 
\begin{eqnarray}
&<&0^{+}\mid T[\widehat{\psi }(x)\widehat{\overline{\psi }}(y)\widehat{C}%
^b(x)\widehat{\overline{C}}^a(z)]\mid 0^{-}>  \nonumber \\
&=&\int d^4y^{\prime }d^4z^{\prime }\gamma ^{ba^{\prime }}(x,y^{\prime
},z^{\prime })S_F(y^{\prime }-y)\Delta ^{a^{\prime }a}(z^{\prime }-z)
\end{eqnarray}
where 
\begin{equation}
\gamma ^{ba^{\prime }}(x,y,z^{\prime })=\int d^4x^{\prime }d^4u^{\prime
}S_F(x-x^{\prime })\Gamma ^{b^{\prime }a^{\prime }}(x^{\prime },y,u^{\prime
},z^{\prime })\Delta ^{b^{\prime }b}(u^{\prime }-x)
\end{equation}
In the above, $S_F(x-x^{\prime }),D_{\mu \nu }^{ab}(x-x^{\prime })$ and $%
\Delta ^{ab}(x-x^{\prime })$ are the quark, gluon and ghost particle
propagators respectively, $\Gamma ^{a\mu }(x,y,z)$ denotes the three-line
quark-gluon proper vertex and $\Gamma ^{ab}(x,y,z,u)$ designates the
four-line quark-ghost vertex. Upon substituting Eqs.(4.2), (4.3) and (4.5)
in Eq.(4.1) and transforming Eq.(4.1) to the momentum space, the Ward
identity will be written as 
\begin{eqnarray}
&&S_F(p)\Gamma ^{b\nu }(p,q,k)S_F(q)k^\mu D_{\mu \nu }^{ab}(k)  \nonumber \\
&=&\xi g[\gamma ^b(p,q,k)S_F(q)-S_F(p)\overline{\gamma }^b(p,q,k)]\Delta
^{ab}(k)
\end{eqnarray}
where 
\begin{equation}
\gamma ^b(p,q,k)=T^c\gamma ^{bc}(p,q,k)
\end{equation}
\begin{equation}
\overline{\gamma }^b(p,q,k)=\gamma ^{bc}(p,q,k)T^c
\end{equation}
Employing the expressions denoted in Eqs.(2.4) and (2.11) and operating on
the both sides of Eq.(4.7) by $S_F^{-1}(p)$ from the left and by $%
S_F^{-1}(q) $ from the right. we arrive at $^{}$ 
\begin{eqnarray}
k^\mu \Gamma _\mu ^a(p,q,k) &=&g[1+\omega (k^2)]^{-1}[S_F^{-1}(p)\gamma
^a(p,q,k)  \nonumber \\
&&-\overline{\gamma }^a(p,q,k)S_F^{-1}(q)]
\end{eqnarray}
This just is the Ward identity satisfied by the quark-gluon vertex$^{32}$.
Considering the energy-momentum conservation and introducing new vertex
functions $\Lambda _\mu ^a(p,q)$ and $\chi ^a(p,q)$ which are defined by 
\begin{equation}
\Gamma _\mu ^a(p,q,k)=(2\pi )^4\delta ^4(p-q-k)g\Lambda _\mu ^a(p,q)
\end{equation}
\begin{equation}
\gamma ^a(p,q,k)=(2\pi )^4\delta ^4(p-q-k)[1+\omega (k^2)]\chi ^a(p,q)
\end{equation}
\begin{equation}
\overline{\gamma }^a(p,q,k)=(2\pi )^4\delta ^4(p-q-k)[1+\omega (k^2)]%
\overline{\chi }^a(p,q)
\end{equation}
the identity in Eq.(4.10) can be represented in the form 
\begin{equation}
(p-q)^\mu \Lambda _\mu ^a(p,q)=S_F^{-1}(p)\chi ^a(p,q)-\overline{\chi }%
^a(p,q)S_F^{-1}(q)
\end{equation}
In the lowest order approximation, the above identity is clearly satisfied
as long as we notice 
\begin{eqnarray}
\Lambda _\mu ^{(0)a}(p,q) &=&i\gamma _\mu T^a \\
\chi ^{(0)a}(p,q) &=&\overline{\chi }^{(0)a}(p,q)=iT^a
\end{eqnarray}
The latter result may easily be derived from the definitions given in
Eqs.(4.4), (4.6), (4.8), (4.9), (4.12) and (4.13) by using the lowest order
expression 
\begin{eqnarray}
\Gamma ^{ab}(x,y,z,u)\approx iS_F^0(x-y)^{-1}\Delta _{ab}^0(z-u)^{-1}
\end{eqnarray}

Now, we are interested in the one-loop approximation of order $g^2$. In this
approximation, the quark-gluon vertex denoted by $\Lambda _\mu ^{(1)a}(p,q)$
is contributed from the two diagrams in Figs.(1h) and (1i) whose expressions
are well known. The quark-ghost vertex functions $\chi ^a(p,q)$ and $%
\overline{\chi }^a(p,q)$ evaluated from Figs.(1j) and (1k) are shown in the
following. 
\begin{equation}
\chi ^{(1)a}(p,q)=iT^aI(p,q)
\end{equation}
where 
\begin{equation}
I(p,q)=-\frac 32g^2\int \frac{d^4l}{(2\pi )^4}S_F(l)\gamma ^\mu D_{\mu \nu
}(q-l)(l-p)^\nu \Delta (l-p)
\end{equation}
and 
\begin{equation}
\overline{\chi }^{(1)a}(p,q)=iT^a\overline{I}(p,q)
\end{equation}
where 
\begin{eqnarray}
\overline{I}(p,q) &=&-\frac 32g^2\int \frac{d^4l}{(2\pi )^4}\gamma ^\mu
S_F(l)(l-q)^\nu  \nonumber \\
&&\times D_{\mu \nu }(l-p)\Delta (l-q)
\end{eqnarray}
It is clear that the above functions are logarithmically divergent. Thus, up
to the order $g^2$, we can write 
\begin{equation}
\Lambda _\mu ^{(1)a}(p,q)=iT^a[\gamma _\mu +\Lambda _\mu ^{(1)}(p,q)]
\end{equation}
where we have set $\Lambda _\mu ^a(p,q)=iT^a\Lambda _\mu ^{(1)}(p,q)$, 
\begin{equation}
\chi ^a(p,q)=iT^a[1+I(p,q)]
\end{equation}
and 
\begin{equation}
\overline{\chi }^a(p,q)=iT^a[1+\overline{I}(p,q)]
\end{equation}

~~Upon differentiating the both sides of Eq.(4.14) with respect to $p^\mu $,
setting q=p and substituting the expression of the inverse of quark
propagator 
\begin{equation}
S_F^{-1}(p)={\not p}-m-\Sigma (p)
\end{equation}
in the order of $g^2$, we get 
\begin{equation}
\overline{\Lambda }_\mu (p,p)=-\frac{\partial \Sigma (p)}{\partial p^\mu }
\end{equation}
where 
\begin{eqnarray}
\overline{\Lambda }_\mu (p,p) &=&\Lambda _\mu ^{(1)}(p,p)-\gamma _\mu
I(p,p)-({\not p}-m)\frac{\partial I(p,q)}{\partial p^\mu }\mid _{q=p} 
\nonumber \\
+\frac{\partial \overline{I}(p,q)}{\partial p^\mu } &\mid &_{q=p}({\not p}-m)
\end{eqnarray}
It is emphasized that at one-loop level, the both sides of Eq.(4.26) are of
the order of $g^2$. In the derivation of Eq.(4.26) from Eq.(4.14), the terms
higher than the order $g^2$ have been neglected. The identity in Eq.(4.26)
formally is the same as we met in QED. The vertex $\stackrel{}{\overline{%
\Lambda }_\mu (p,p)}$ may be expressed in the form 
\begin{equation}
\stackrel{}{\overline{\Lambda }_\mu (p,p)}=L\gamma _\mu +\Lambda _\mu ^c(p)
\end{equation}
where L is a divergent constant defined by 
\begin{equation}
\stackrel{}{L=\overline{\Lambda }_\mu (p,p)}\mid _{\not p=\mu }
\end{equation}
and $\Lambda _\mu ^c(p)$ is the finite part of $\overline{\Lambda }_\mu
(p,p) $ satisfying the boundary condition 
\begin{equation}
\Lambda _\mu ^c(p)\mid _{\not p=\mu }=0
\end{equation}
Integrating the identity in Eq.(4.26) over the momentum $p_{\mu \text{ }}$%
and considering the expression in Eq.(4.28), we have

\begin{equation}
\Sigma (p)=A+({\not p}-\mu )[B-C(p^2)]
\end{equation}
where A and B are the divergent constants depending on the renormalization
point $\mu $ which are defined as 
\begin{equation}
A=\Sigma (\mu )
\end{equation}

\begin{equation}
B=-L
\end{equation}
and $C(p^2)$ is a finite function defined by 
\begin{equation}
\int_{p_{0^{}}^\mu }^{p^\mu }dp^\mu \Lambda _\mu ^c(p)=(\not p-\mu )C(p^2)
\end{equation}
with boundary condition 
\begin{equation}
C(p^2)=0
\end{equation}
Clearly, the expression in Eq.(4.31) gives a subtraction version for the
fermion self-energy which is required by the Ward identity and correct at
least in the approximation of the order g$^2$. With this subtraction, the
quark propagator will be renormalized as follows 
\begin{equation}
S_F(p)=\frac{Z_2}{\not p-m_R-\Sigma _R(p)}
\end{equation}
where Z$_{2,}$ m$_R$ and $\Sigma _R(p)$ are the quark propagator
renormalization constant, the renormalized quark mass and the finite
correction of the self-energy respectively. The Z$_2$ and the m$_{R\text{ }}$%
are respectively defined as 
\begin{equation}
Z_2^{-1}=1-B
\end{equation}
and 
\begin{equation}
m_R=Z_m^{-1}m
\end{equation}
where Z$_m$ is the quark mass renormalization constant which is defined by
the following expression 
\begin{equation}
Z_m^{-1}=1+Z_2[Am^{-1}+(1-\mu m^{-1})B]
\end{equation}
\[
\]

\setcounter{section}{5}

\section*{5.Effective Quark Mass}

\setcounter{equation}{0}

~~~Taking the derivative of Eq.(4.38) with respect to $\mu $ and noticing $%
\mu \frac d{d\mu }=\lambda \frac d{d\lambda }$, we get a RGE for the
renormalized quark mass as follows 
\begin{equation}
\lambda \frac d{d\lambda }m_R(\lambda )+\gamma _m(\lambda )m_R(\lambda )=0
\end{equation}
where 
\begin{equation}
\gamma _m(\lambda )=\lim_{\varepsilon \to 0}\mu \frac d{d\mu }\ln Z_m
\end{equation}
is the mass anomalous dimension. Let us concentrate our attention on the
one-loop approximation. The fermion self-energy of the one-loop diagram
shown in Fig.(1l) is of the following regularized form in the n-dimensional
space$^{31,32}$ 
\begin{eqnarray}
\Sigma (p) &=&-\frac{g^2}{12\pi ^2}(4\pi M^2)^\varepsilon \Gamma
(1+\varepsilon )\int_0^1dx\{\frac 1{\varepsilon \Delta (p^2)^\varepsilon }%
\stackrel{}{_{}}  \nonumber \\
&&\ \ \times [2(1-\varepsilon )(1-x)\not p{}-2(2-\varepsilon )m+(1-\xi
)(m-2xp)]  \nonumber \\
&&\ \ -2(1-\xi )(1-x)x^2\frac{p^2{\not p}}{\Delta (p^2)^{1+\varepsilon }}\}
\end{eqnarray}
where 
\begin{equation}
\Delta (p^2)=p^2x(x-1)+m^2x
\end{equation}
Substituting the above expression in Eq.(4.31), one may find 
\begin{eqnarray}
A=\Sigma (p)|_{{\not p}=\mu }= &&-\frac{g^2}{12\pi ^2}(4\pi M^2)^\varepsilon
\Gamma (1+\varepsilon )\int_0^1dx\{\frac 1{\varepsilon \Delta (\mu
^2)\varepsilon }  \nonumber \\
&&\ \ \times [2\mu [1+(\xi -2)x-\varepsilon (1-x)]-(3+\xi -2\varepsilon )m] 
\nonumber \\
&&\ \ -2(1-\xi )(1-x)x^2\frac{\mu ^3}{\Delta (\mu ^2)^{1+\varepsilon }}\}
\end{eqnarray}
and 
\begin{eqnarray}
B &=&[\Sigma (p)-A]({p}-\mu )^{-1}|_{{\not p}=\mu }  \nonumber \\
\ &=&-\frac{g^2}{12\pi ^2}(4\pi M^2)^\varepsilon \Gamma (1+\varepsilon
)\int_0^1dx\{\frac 1{\varepsilon \Delta (\mu ^2)\varepsilon }  \nonumber \\
&&\ \ \times [2(1-\varepsilon )(1-x)-2(1-\xi )x]+\frac{2\mu ^2}{\Delta (\mu
^2)^{1+\varepsilon }}x(x-1)  \nonumber \\
&&\ \ \times [2(1-\varepsilon )(x-1)+5(1-\xi )x+\frac m\mu (3+\xi
-2\varepsilon )]  \nonumber \\
&&\ \ -\frac{4\mu ^4}{\Delta (\mu ^2)^{2+\varepsilon }}(1-\xi
)(1+\varepsilon )(x-1)^2x^3\}
\end{eqnarray}
where 
\begin{equation}
\Delta (\mu ^2)=x[(x-1)\mu ^2+m^2]
\end{equation}
By making use of the renormalization constants defined in Eqs.(4.39) and
(4.37) and the constants given in Eqs.(5.5) and (5.6), in the order of $g^2$%
, it is not difficult to derive the anomalous dimension defined in Eq.(5.2)
for i-th quark as shown in the following 
\begin{equation}
\gamma _m^{(i)}(\lambda )=\frac{\alpha _R(\lambda )}\pi F_m^{(i)}(\lambda )
\end{equation}
where 
\begin{eqnarray}
F_m^{(i)}(\lambda ) &=&\frac{2\xi }{3c_i}\lambda +2[3+2\xi -3(1+\xi )\frac{%
c_i}\lambda  \nonumber \\
&&\ \ +\frac{2\xi c_i^2}{\lambda ^2}]-\frac{4(1+\xi )\lambda }{c_i+\lambda }-%
\frac{2c_i^2}{\lambda ^2}[3+\xi  \nonumber \\
&&\ \ -3(1+\xi )\frac{c_i}\lambda +\frac{2\xi c_i^2}{\lambda ^2}]\ln \left| 
\frac{c_i^2}{c_i^2-\lambda ^2}\right|
\end{eqnarray}
in which $c_i=\frac{m_R^i}{\mu _0}$. The $\mu _0$ may be chosen as stated in
Sect.3. As seen from Eq.(5.9), the anomalous dimension and hence the
effective mass are gauge-dependent at one-loop level. In the following. we
are only interested in the result given in the Landau gauge which was
regarded as preferred gauge in the previous literature$^{14}$. In this
gauge, 
\begin{eqnarray}
F_m^{(i)}(\lambda ) &=&6-\frac{6c_i}\lambda -\frac{4\lambda }{\lambda +c_i}+%
\frac{6c_i^2}{\lambda ^2}(1-\frac{c_i}\lambda )  \nonumber \\
&&\ \ \times \ln \left| 1-\frac{\lambda ^2}{c_i^2}\right|
\end{eqnarray}

Substituting Eqs.(5.8) and (5.9) in Eq.(5.1) and solving the equation, the
effective mass is found to be 
\begin{equation}
m_R^{(i)}(\lambda )=m_R^{(i)}e^{-S^{(i)}(\lambda )}
\end{equation}
where 
\begin{equation}
S^{(i)}(\lambda )=\frac 1\pi \int_1^\lambda \frac{d\lambda }\lambda \alpha
_R(\lambda )F_m^{(i)}(\lambda )
\end{equation}
here the $\alpha _R(\lambda )$ was shown in Eq.(3.10). If we take $\alpha
_R(\lambda )\approx \alpha _R^0$ and work in the Landau gauge, we get 
\begin{eqnarray}
S^{(i)}(\lambda ) &=&\frac{\alpha _R^0}\pi \{2c_i(\frac 1\lambda -1)+(\frac{%
2c_i^3}{\lambda ^3}-\frac{3c_i^2}{\lambda ^2}+1]  \nonumber \\
&&\ \times \ln \left| 1-\frac{\lambda ^2}{c_i^2}\right|
-(2c_i^3-3c_i^2+1)\ln \left| 1-\frac 1{c_i^2}\right| \}
\end{eqnarray}
If we take $c_i=1$, namely, set all masses to be equal and choose $\mu
_0=m_R^i$, Eq.(5.13) will be reduced to 
\begin{equation}
S^{(i)}(\lambda )=\frac{\alpha _R^0}\pi \frac{(1-\lambda )}\lambda
\{2+[\frac 2{\lambda ^2}-\frac{1+\lambda }\lambda ]\ln \left| 1-\lambda
^2\right| \}
\end{equation}
where 
\begin{equation}
\ln \left| 1-\lambda ^2\right| =\cases{ 2[\ln (1+\lambda )-\tanh^{-1}\lambda
],&if$ \lambda <1;$ \cr 2[\ln (1+\lambda )-\coth^{-1}\lambda ],&if$ \lambda
>1$\cr}
\end{equation}
The behavior of effective masses is illustrated in Fig.(4). The results in
Fig.(4) are given by taking $N_f=4$ and all the quark masses to be equal.
The solid curve and the dashed one represent the effective masses calculated
respectively by taking the coupling constant to be the running one and a
fixed value $\alpha _R=0.2$. The common feature of these effective masses is
as follows. For each curve, there is a maximum at $\lambda \approx 1.54$.
When $\lambda $ tends to infinity, the masses fall down to zero, exhibiting
the well-known asymptotically free behavior. For the $\lambda $ lying in the
region [0,1), the mass is less than the maximum and behaves almost as a
constant. However, in the region of $\lambda $ near zero, as pointed out in
Sect.3, the QCD perturbative results are no longer valid.

\setcounter{section}{6}

\section*{6.Comments and Conclusions}

\setcounter{equation}{0}

~~~In this paper, the QCD one-loop renormalization has been restudied in the
GMS scheme. It was shown that the GMS scheme allows us not only to consider
the mass effect on the renormalization, but also to directly relate the
renormalization scale $\mu $ to the momentum p. The effective coupling
constant and the effective quark masses obtained in this scheme get a
noteworthy improvement near the heavy quark threshold in comparison with
those given previously in the MS scheme. The effective coupling constant and
the effective quark mass presented in the MS scheme now appear as
approximate results given in the large momentum limit. It is noted here that
even in this limit, the effective coupling constant and the effective mass
can only have unique forms. It is impossible to result in a difference
between the $MS$ and $\overline{MS}$ schemes$^{7,8,16}$ because an effective
quantity should be the solution of its RGE whose form is uniquely determined
by the anomalous dimension and the boundary condition. As emphasized before,
the correct procedure of evaluating the anomalous dimension is the use of
the regularized renormalization constant and the limit $\varepsilon
\rightarrow 0$ should be taken after differentiation with respect to the
renormalization point. By this procedure, the factor $(4\pi M)^\varepsilon
\Gamma (1+\varepsilon )$ appearing in the n-dimensional Feynman integrals
can only come to unity. It is not possible to yield the unphysical terms $%
\ln 4\pi -\gamma $ in a renormalized quantity.

In this paper, the effective coupling constant given in the one-loop
approximation is calculated by employing the renormalization constants given
by the subtractions of the gluon and ghost particle self-energies and the
ghost vertex correction. It is emphasized that the renormalization constant $%
Z_2$ is obtained on the basis of the Ward identity obeyed by the gluon
self-energy and therefore is faithful. The renormalization constants $%
\widetilde{Z}_1$ and $\widetilde{Z}_3$ derived respectively from the
subtractions of the ghost vertex correction and the ghost particle
self-energy, as mentioned in Sect.4, give the anomalous dimensions which are
scheme-independent due to that there is no quark mass to appear in the
vertex and the self-energy. Therefore, the anomalous dimension $\gamma _g$
computed from these renormalization constants is definite, no any
uncertainty. Particularly, the correctness of the anomalous dimension shown
in Eqs.(3.6)-(3.8) is confirmed by the previous result presented in Refs.(9)
and (16) where the $\gamma _g$ was determined by the subtraction of the
quark and gluon self-energies and the quark-gluon vertex at a space-like
point. The expression of the $\beta $-function given in the space-like point
is $^{3,10}$%
\begin{eqnarray}
\beta &=&-g\gamma _g=-\frac{g^3}{16\pi ^2}\{11-\frac 23\sum_{i=1}^{N_f}[1-%
\frac{6c_i^2}{\lambda ^2}  \nonumber \\
&&\ +\frac{12c_i^4}{\lambda ^3(\lambda ^2+4c_i^2)^{\frac 12}}\ln \frac{%
(\lambda ^2+4c_i^2)^{\frac 12}+\lambda }{(\lambda ^2+4c_i^2)^{\frac
12}-\lambda }]\}
\end{eqnarray}
where we have set $\mu =\mu _0\lambda $ and $c_i=\frac{m_R}{\mu _0}$. The
above expression may be directly written out from Eqs.(3.6)- (3.8) by the
transformation $c_i^2\to -c_i^2$, corresponding to $\mu ^2\to -\mu ^2$. The
effective coupling constant given by the $\beta $-function in Eq.(6.1) still
exhibits the property of asymptotic freedom when $\lambda \to \infty $; but,
it has a Landau singularity in the region of large distance, different from
the result given in the GMS scheme. In Fig.(2), the Landau singularity
occurs at $\lambda \approx 0.4$.

It is stressed again that the effective quark masses shown in Sect.5 are
obtained based on the subtraction written in Eq.(4.31) which is derived from
the Ward identity respected by the quark one-loop self-energy. These
effective masses are different from those given in the previous works$%
^{3,10} $. For example, in Ref.(3), the mass anomalous dimensions derived in
the Landau gauge is of the form 
\begin{equation}
\gamma _m^{(i)}=\frac{g^2}{2\pi ^2}[1-\frac{m_i^2}{\mu ^2}\ln (1+\frac{\mu ^2%
}{m_i^2})]
\end{equation}
which is manifestly different from that formulated in Eqs.(5.8) and (5.10).
The result shown above , actually, was obtained from another subtraction. If
we start from the expression for the quark self-energy$^{29}$ 
\begin{equation}
\Sigma (p)=A(p^2)P+B(p^2)m
\end{equation}
and subtract at the space-like point $p^2=-\mu ^2$, the quark propagator and
mass renormalization constants will be defined by 
\begin{equation}
Z_2^{-1}=1-A(-\mu ^2)
\end{equation}
and 
\begin{equation}
Z_m^{-1}=Z_2[1+B(-\mu ^2)]
\end{equation}
In the approximation of order $g^2$ for the one-loop diagram, the constant $%
Z_m$ defined in Eq.(6.5) will give rise to the result in Eq.(6.2). Since the
subtraction above is not compatible with the Ward identity, the effective
quark masses obtained in this subtraction can not be viewed as reasonable
results. The correct result of the effective quark mass given by the
subtraction at space-like may be obtained from Eqs.(5.11)-(5.15) through the
transformation $\lambda \to i\lambda $. In this case, as we see, the
effective mass becomes complex. If we require the effective mass to be real,
the subtraction at space-like point should be excluded. However, this does
not mean that the space-like momentum subtraction is useless. The
subtractions at time-like point and at space-like point probably suit to
different processes in which the interactions are of different behaviors.
For example, for the t- channel fermion-antifermion scattering, the momentum
on the intermediate boson line is space-like; while, for the s-channel
scattering, the corresponding momentum is time-like. It seems that the boson
self-energy should be subtracted at the space-like point for the former
process and at time-like point for the latter process. This problem is, we
think, worthy to pursue in future investigations.

\section{Acknowledgment}

The authors would like to thank Professor Shi-Shu Wu for useful discussions.
This work was supported in part by National Natural Science Foundation of
China.

\section{FIGURE CAPTIONS}

Fig.(1) The one-loop diagrams.

Fig.(2) The one-loop effective coupling constants given by taking $%
\alpha^0_R=0.2, N_f=4$ and equal mass of quarks. The solid curve represents
the result obtained at time-like subtraction point. The dashed curve
represents the one given at space-like subtraction point. The dotted curve
shows the MS scheme result.

Fig.(3) The one-loop effective coupling constants evaluated for the
different numbers of flavors $N_f=1,3,6$ and for unequal quark masses. The
masses of quarks of different flavors are taken to be the values as written
in Sect.(3).

Fig.(4) The one-loop effective quark masses obtained by taking $N_f=4$ and
equal mass. The solid curve represents the effective mass obtained by using
the running coupling constant. The dashed curve represents the one given by
taking the coupling constant to be a constant $\alpha^0_R=0.2$.


\begin{thebibliography}{99}
\bibitem{1}  H.D.Politzer. Phys.Rev.Lett.30, 1346 (1973).

\bibitem{2}  D.J.Gross and F.Wilczek, Phys.Rev.Lett.30, 1343 (1973);
Phys.Rev. D8, 3633 (1973).

\bibitem{3}  H.Georgi and H.D.Politzer, Phys.Rev.D14, 1829 (1976).

\bibitem{4}  E.G.Floratos, D.A.Ross and C.T.Sachrajda, Nucl.Phys.B129, 66
(1977); B139, 54 (1978); Phys.Lett.80B, 269 (1979).

\bibitem{5}  G.Altarelli, R.K.Ellis and G.Martinelli, Nucl.Phys.B143, 521
(1978).

\bibitem{6}  G. t' Hooft, Nucl. Phys.B61, 455 (1973).

\bibitem{7}  W.A.Bardeen, A.J.Buras, D.W.Duke and T.Muta, Phys.Rev.D18, 3998
(1978); W.A.Bardeen and R. A.J.Buras, Phys.Rev.D20, 166 (1979).

\bibitem{8}  W. Celmaster and R. J.Gonsalves, Phys.Rev.Lett.42, 1435 (1979);
Phys.Rev.D20, 1420 (1979);

W.Celmaster and D.Sivers, Phys.Rev.D23, 227 (1981).

\bibitem{9}  E.Braaten and J.P.Leveille, Phys.Rev.D24, 1369 (1981).

\bibitem{10}  S.N.Gupta and S.F.Radford, Phys.Rev. D25, 2690 (1982).

\bibitem{11}  D.M.Stevenson, Phys.Rev.D23, 2916 (1981).

\bibitem{12}  D.Espriu and R.Tarrach,Phys.Lett.102B, 163 (1981);
Phys.Rev.D25, 1073 (1982).

\bibitem{13}  G.Grunberg, Phys. Rev. D29, 2315 (1984).

\bibitem{14}  A. Dhar, Phys.Lett. B128, 407 (1983); A. Dhar and V. Gupta,
Phys.Rev.D29, 2822 (1984).

\bibitem{15}  G.B.West, Phys.Rev.D30, 1349 (1984).

\bibitem{16}  D.A.Raczka and R.Raczka, Phys.Rev. D39, 643 (1989); D40, 878
(1989).

\bibitem{17}  D.J.Broadhurst, N.Gray, and K.Schilcher, Z.Phys.C52, 111
(1991).

\bibitem{18}  J.Fleischer and O.V.Tarasov, Phys.Lett.B283, 129 (1992).

\bibitem{19}  M. Beneke, Phys. Lett. B307, 154 (1993).

\bibitem{20}  L. E. Adam and K. G. Chetyrkin, Phys.Lett. B329, 129 (1994).

\bibitem{21}  P. A. Raczka, Z. Phys. C65, 481 (1995); P. A. Raczka and A.
Szymacha, Z. Phys. C70, 125 (1996).

\bibitem{22}  S. Sint, Nucl.Phys. B45, 416 (1995).

\bibitem{23}  J. A. Gracey, Nucl. Phys. Proc. Suppl. 51C, 24 (1996).

\bibitem{24}  N.N.Bogoliubov and D.V.Shirkow, Introduction to the Theory of
Quantized Fields, Wiley-Interscience, New York (1959).

\bibitem{25}  M.Gell-Manm and F.E.Low, Phys.Rev.95, 1300 (1954).

\bibitem{26}  C.G.Callan, Phys.Rev.D2, 1541 (1970).

\bibitem{27}  K.Symanzik, Commun. Math. Phys.18, 227 (1970).

\bibitem{28}  S.Weinberg, Phys.Rev.D8, 3497 (1973).

\bibitem{29}  J.C.Collins and A.J.Macfarlane, Phys.Rev.D10, 1201 (1974).

\bibitem{30}  J.C.Taylor, Nucl. Phys.B33, 436 (1971); A.A.Slavnov, Theor.
Math.Phys.10, 99 (1972).

\bibitem{31}  C. Itzykson and J-B. Zuber, Quantum Field Theory, McGraw-Hill,
New York (1980).

\bibitem{32}  Bing-Lin Young, Introduction to Quantum Field Theories,
Science Presss, Beijing (1987).
\end{thebibliography}
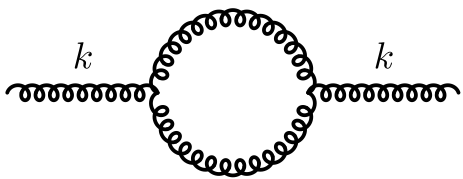
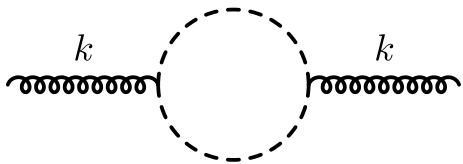
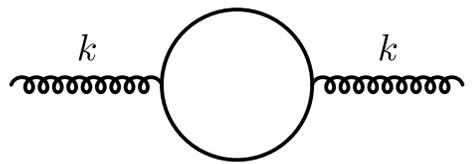
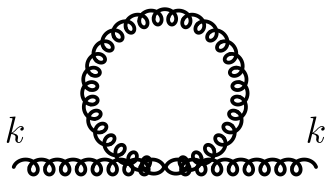
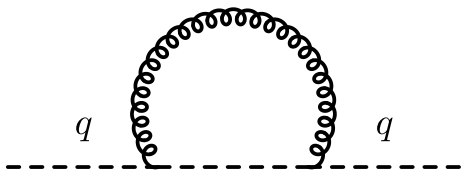
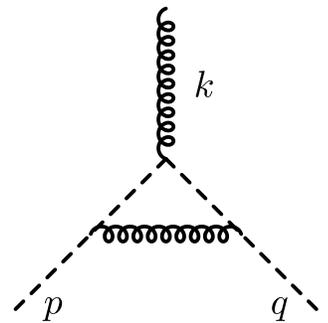
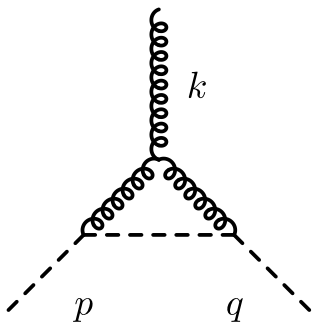
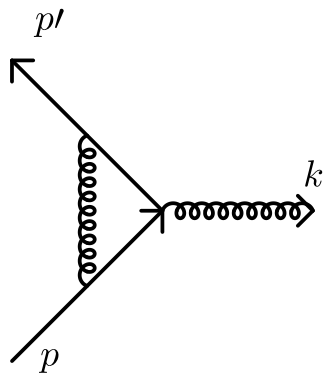
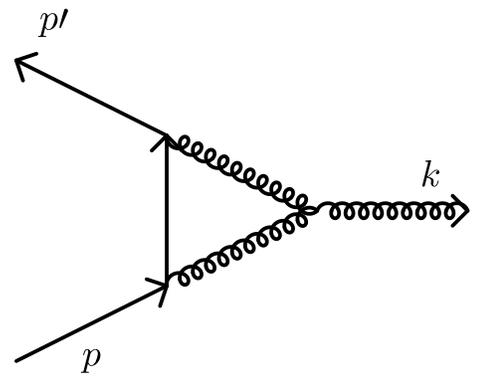
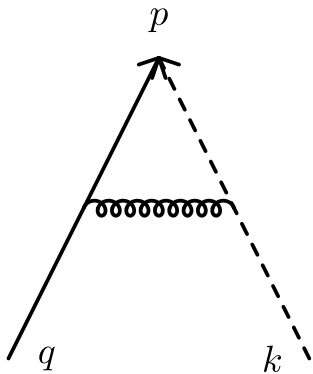
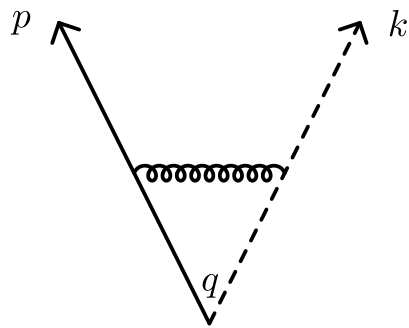
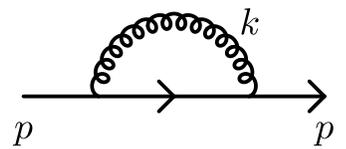
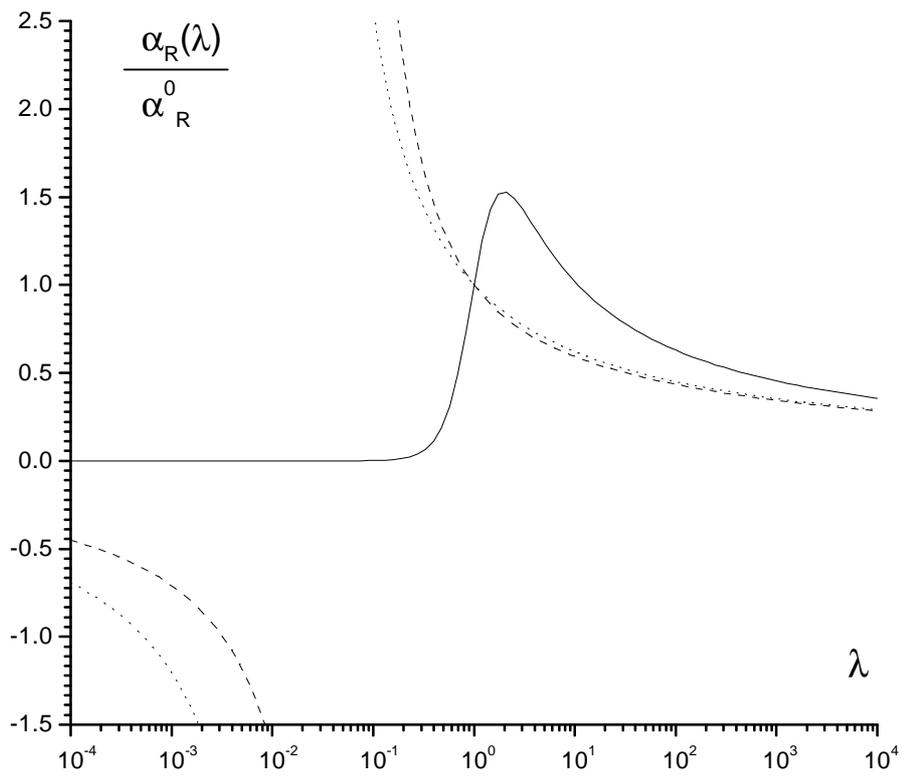
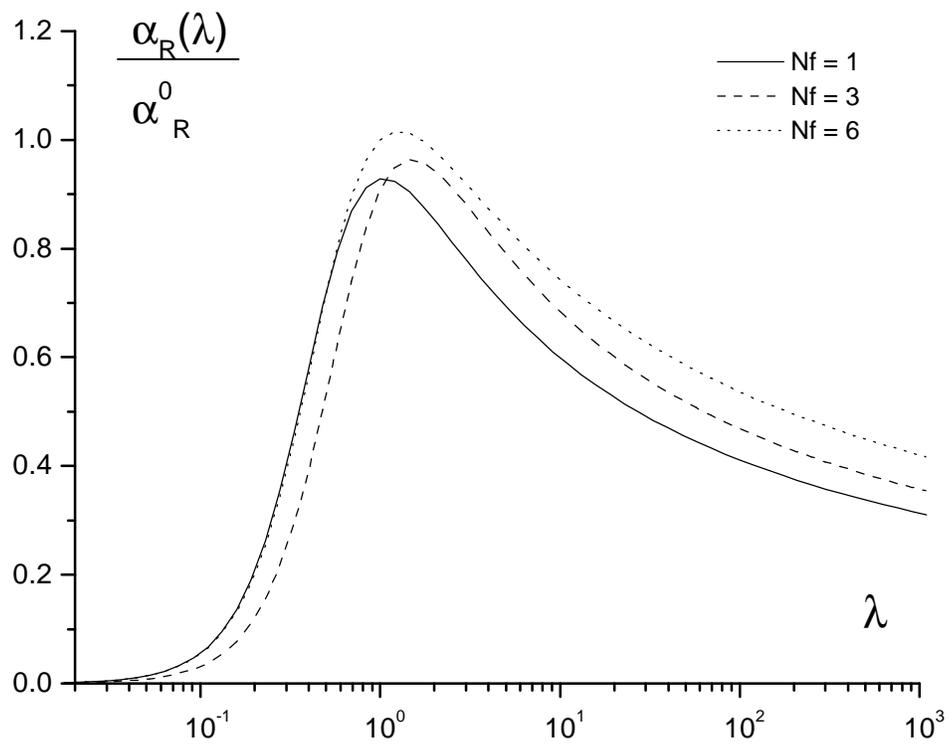
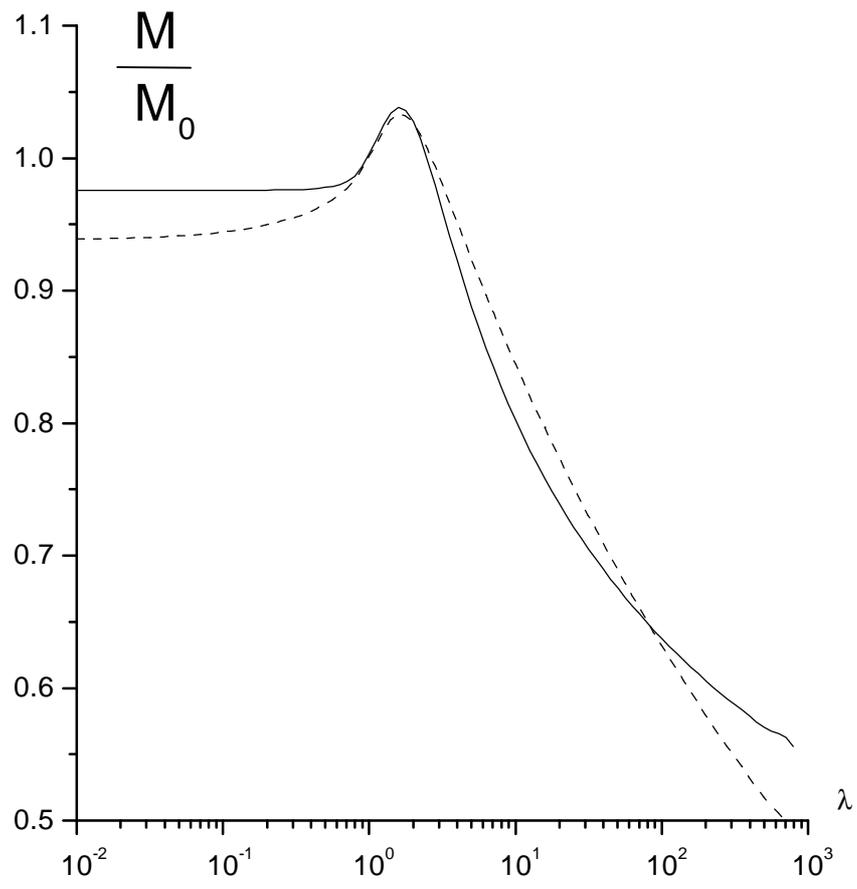
\end{document}